\documentclass[signals,article,submit,pdftex,moreauthors]{Definitions/mdpi} 
\preto{\abstractkeywords}{\nolinenumbers}

\firstpage{1} 
\pubvolume{1}
\issuenum{1}
\articlenumber{0}
\pubyear{2022}
\copyrightyear{2022}
\datereceived{} 
\dateaccepted{} 
\datepublished{} 
\hreflink{https://doi.org/} 

\usepackage{graphicx}
\usepackage{multirow}
\usepackage{booktabs}
\usepackage{url}

\Title{Personalized PPG Normalization based on Subject Heartbeat in Resting State Condition}

\TitleCitation{Personalized PPG Normalization based on Subject Heartbeat in Resting State Condition}

\Author{Francesca Gasparini $^{1}$*\orcidA{}, 
 Alessandra Grossi $^{1}$\orcidB{}, Marta Giltri  $^{1}$\orcidC{} and Stefania Bandini $^{2,}$ $^{1}$\orcidD{}}

\AuthorNames{Francesca Gasparini, Marta Giltri, Alessandra Grossi and Stefania Bandini}

\AuthorCitation{Gasparini, F.; Giltri, M.; Grossi, A.; Bandini, S.}

\address{%
$^{1}$ \quad Department of Informatics, Systems and Communication, University of Milano-Bicocca, Milan, Italy; francesca.gasparini@unimib.it (F.G.),  a.grossi6@campus.unimib.it (A.G.), m.giltri@campus.unimib.it (M.G.), stefania.bandini@unimib.it (S.B.)\\
$^{2}$ \quad RCAST - Research Center for Advanced Science \& Technology, The University of Tokyo, Tokyo, Japan}

\corres{Correspondence: francesca.gasparini@unimib.it}

\abstract{Physiological responses are nowadays widely used to recognize the affective state of subjects in real-life scenarios. However, these data are intrinsically subject-dependent, making machine learning techniques for data classification not easily applicable due to inter-subject variability.
In this work, the reduction of inter-subject heterogeneity is considered in the case of PhotoPlethysmoGraphy (PPG), which is successfully used to detect stress and evaluate experienced cognitive load. 
To face the inter-subject heterogeneity, a novel personalized PPG normalization is here proposed. A subject-normalized discrete domain where the PPG signals are properly re-scaled is introduced, considering the subject's heartbeat frequency in resting state conditions. The effectiveness of the proposed normalization is evaluated in comparison with other normalization procedures in a binary classification task, where cognitive load and relaxing state are considered. The results obtained on two different datasets available in the literature confirm that applying the proposed normalization strategy permits to increase classification performance.}


\begin{document}

\section{Introduction}

In recent years, sensor technology has improved significantly and wearable devices have become increasingly popular, allowing to easily register subjects' physiological responses during their daily activities \cite{can2019stress, han2020objective}. 
Physiological signals are successfully used to measure arousal. Arousal is a human uncontrolled reaction, related to attention and cognitive alertness, activated by stimuli that require high psycho-physical engagement, and thus activated in particular during cognitive tasks and stressful conditions.
Although it is proved that sometimes stress can have a positive effect on a person by improving his/her alertness state or his/her ability to react \cite{choi2011development}, it is also proved that a high and continuous level of stress or cognitive load can affect the physical and mental subject's well-being. Illnesses like depression, anxiety and sleep disorders are, indeed, often due to excessive stress or workload \cite{umematsu2019improving}. 

In view of its importance, the automatic recognition of stress and excessive cognitive load has recently become an object of studies and researches, even in different application areas. For instance, systems able to recognize emotion and, above all, stress experienced by subjects can be used in working or academic environment \cite{duran2021electronic} in order to monitor and identify the emotional state of employees or students. In this regard, it is indeed proved that a high level of stress or cognitive load due to excessive workload can increase the level of fatigue, decrease the subject's working capability and, consequently, bring physical and mental illness that can lead to absence from the workplace \cite{setz2009discriminating}. Similarly, automatic stress recognition systems can be used in the context of vehicle driving for the detection of excessive mental fatigue states that can reduce a person's driving skills \cite{saeed2017deep}.      
Finally, algorithms of stress detection can be also used in recreational areas for the development of systems able to modify their parameters based on the user's emotional state. Concrete examples concern music-retrieval systems able to interact with the user of a music playlist using both external inputs and his physiological signals \cite{bandini2019personalized} or videogames in which some internal game parameters, like the difficulty, are set on the base of player's emotions and stressing level \cite{vachiratamporn2014implementation}. Moreover, systems able to recognize the subject's emotional state can also be involved in the medical area. For instance, emotion recognition systems can be used to monitor the health state of convalescent patients \cite{choi2011development} or to help elderly subjects during their daily activities \cite{fonseca2019monitoring, bartolome2020arousal}. 
In all these contexts, the development of systems able to recognize, interpret and simulate human affect can be seen as a necessary step to make technologies user-friendly and able to interact actively with people. 

Despite the progress in sensor technology and the relative simplicity of acquiring physiological signals from the human body, there are still some critical issues that must be addressed to fully exploit the potential that the analysis of physiological signals can offer.
Although it is always easier to acquire data, which can overcome the low cardinality of datasets, the application of machine learning techniques is still limited by inter-subject heterogeneity.
Even in the same resting condition, without external stimuli, physiological signals appear significantly subject dependent.

In this work, the reduction of inter-subject heterogeneity is faced in the case of one of the most used physiological signal acquired by wearable devices: the heartbeat, mainly detected through PhotoPlethysmoGraphy (PPG) \cite{castaneda2018review}. The PPG signal is one of the most used signals to measure arousal \cite{lee2019fast,ayata2018emotion}, and consequently to detect stress and evaluate the experienced cognitive load \cite{xuan2020assessing,kalra2020mental} . 
The PPG signal of each subject appears different, both in terms of amplitude and beat frequency.
Regarding the amplitude, differences can be due to the subjects' skin characteristics or to different sensor adherence during the acquisition phase.
Concerning the diversity in terms of heartbeat, according to the American Heart Association the heart rate frequency of a resting adult can vary in a range between 60 and 100 beats per minute, and it depends on many different factors both personal (like age, sex, ethnicity, sports ability, diet, illnesses, prescribed medications, etc.) and environmental (humidity, temperature, etc.) \cite{avram2019real}. 
Normalization procedures based on data rescaling are often used \cite{goh2020robust} to overcome amplitude variability within subjects. The main strategies adopted involve rescaling to the range $[0, 1]$ \cite{lee2019wearable}, normalizing by dividing by the maximum value of the signal \cite{nemati2016monitoring} and applying Z-score \cite{costadopoulos2019using}.
All these methods, however, do not take into account the effective differences in the subjects' heartbeat, which are not only related to amplitude, but also to frequency.

The aim of this work is to solve this inter-subject variability,  adopting a novel personalized PPG normalization based on the heartbeat of the subject in a resting state condition. 
To validate the normalization procedure here presented, PPG data belonging to two different datasets are considered: the \textit{CLAWDAS} (Cognitive Load and Affective Walkability in Different Age Subjects) dataset, partially introduced in \cite{gaspariniAge} and in \cite{gaspariniDeep} and the \textit{CLAS} (Cognitive Load, Affect and Stress recognition) dataset, available in the literature \cite{markova2019clas}.

The proposed normalization is analyzed considering a binary classification task to discriminate cognitive load form relaxing state and compared with normalization strategies adopted in the literature.

The paper is organized as follows. In Section \ref{sec:dataset}, the two considered datasets are described, while the preliminary signal preprocessing strategies applied to each of them are reported in Section \ref{sec:preprocessing}. The novel personalized PPG normalization strategy, based on subject resting state heartbeat is presented in Section \ref{sec:Normalization}. Extracted features, classification strategies and the adopted cross validation approach are reported in Section \ref{sec:classificationSetting}.
The comparison of classification performances on the two datasets, obtained with different normalization strategies, is then reported and analyzed in Section \ref{sec:results}. Finally, conclusions are drawn in the last Section.


\section{Dataset description}
\label{sec:dataset}
\textit{CLAS}  \cite{markova2019clas} and \textit{CLAWDAS} \cite{gaspariniAge, gaspariniDeep} datasets are here considered to validate the proposed normalization procedure.

In the CLAS dataset, the physiological signals of 60 healthy volunteers (mostly students between 20 and 27 years old, 17 women) were acquired while they were performing interactive or perceptive tasks. In particular, the interactive tasks were introduced to evaluate the level of concentration and the cognitive capacity of different individuals by solving Math Problems, Logic Problems and Stroop Tests. In the perceptive tasks, different emotions have been elicited in the participants by images and video selected from the DEAP dataset \cite{koelstra2011deap}.
During the whole experiment, three types of physiological signals were simultaneously recorded by means of Shimmer sensors (www.shimmersensing.com): Electrocardiography (ECG), Plethysmography (PPG) and ElectroDermal Activity (EDA). The signals were acquired with a sampling rate of 256 Hz and a resolution of 16-bits per sample. In addiction, for each subject, 3D accelerometer data and metadata were also collected.

For the purpose of this work, only PPG signals collected during the interactive tasks are considered. For each participant, this phase of the experiment is characterized by the following steps:
\begin{itemize}
    \item \textit{1 minutes} of \textbf{Baseline} in resting state condition
    \item \textit{3 minutes} of \textbf{Math problems} in which the participant solves different simple mathematical problems (MP) in a limited interval of time.
    \item \textit{30 seconds} of \textbf{Neutral State} (NS), in which neutral audio-visual stimuli are displayed
    \item \textit{3 minutes} of \textbf{Stroop tests} (ST)  where the user is expected to match correctly the color of the text with the meaning of the word, having a strict time constraint for each assignment.
    \item \textit{30 seconds} of \textbf{Neutral State} (NS), in which neutral audio-visual stimuli are displayed
    \item \textit{5 minutes} of \textbf{Logic problems} (LP) consisting of several simple logical problems often used during the IQ tests.   
    \item \textit{30 seconds} of \textbf{Neutral State} (NS), in which neutral audio-visual stimuli are displayed
\end{itemize}
Moreover the 3 \textbf{Neutral State} repetitions of \textit{30 seconds} for each subject, in the Picture test of the perceptive session are also considered. For further details of this dataset, please refer to \cite{markova2019clas}.
\newline\newline
\begin{figure}[!tb]
   \begin{minipage}{0.70\textwidth}
     \centering
     \includegraphics[width=0.9\linewidth]{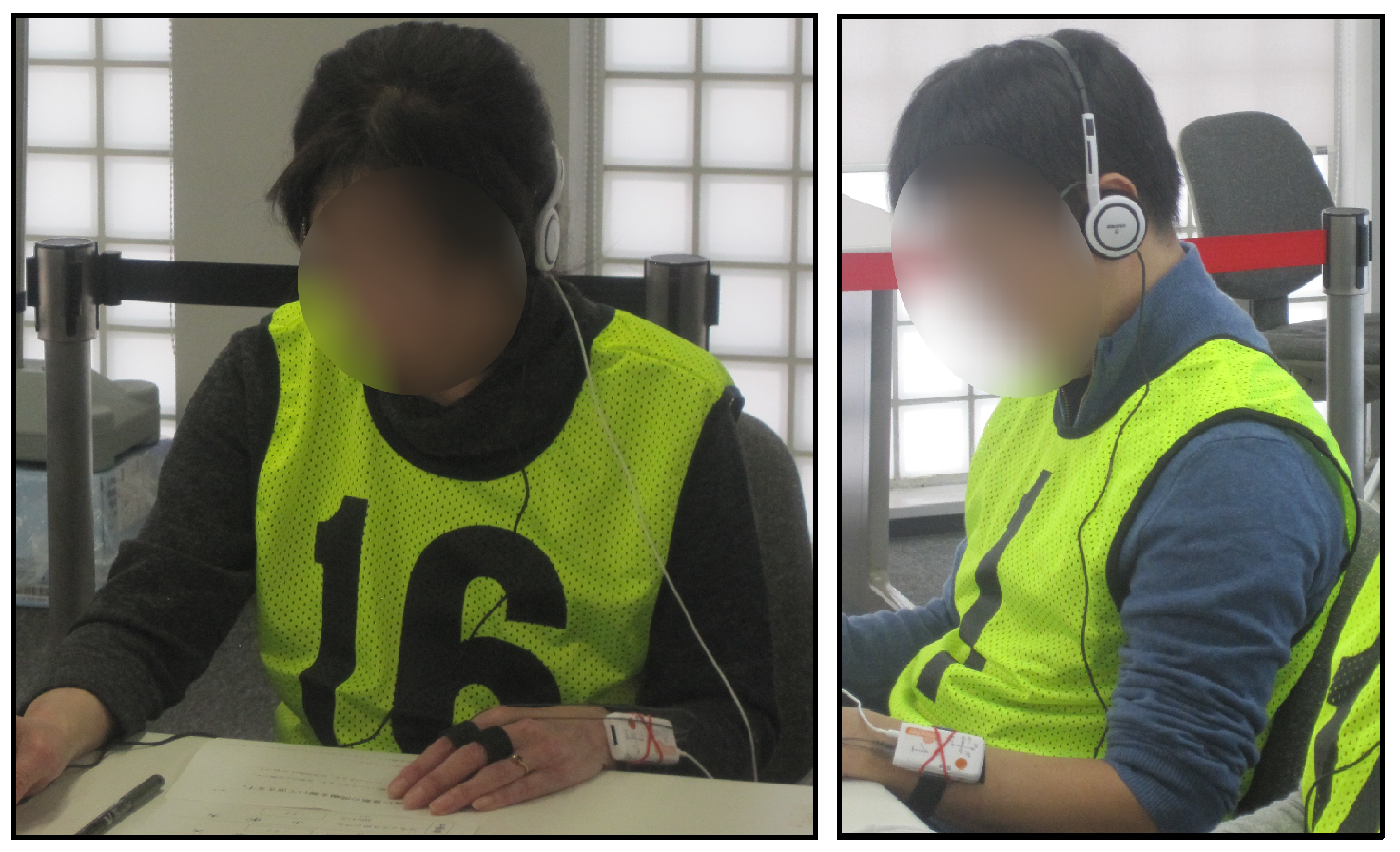}
     \caption{Example of signal acquisition in CLAWDAS dataset:  \\ Left: Math Calculation, Right: Relaxing Audio Listening }
     \label{Fig:example_acquisition}
   \end{minipage}\hfill
   \begin{minipage}{0.30\textwidth}
     \centering
     \includegraphics[width=0.9\linewidth]{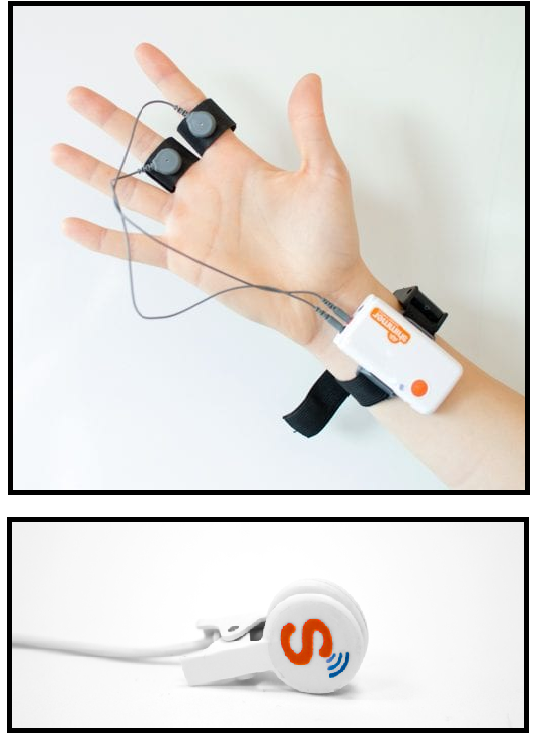}
     \caption{Sensors used to collect physiological data in the CLAWDAS dataset}
     \label{Fig:sensors}
   \end{minipage}
\end{figure}
CLAWDAS is a dataset collected in a  controlled laboratory environment at the Research Center for Advanced Science and Technology (RCAST) at The University of Tokyo. The experiments performed at RCAST are focused on finding differences in physiological responses related to different ages of the subjects, involved in several tasks, from cognitive, to listening and walking ones. The experiments involved two different groups of subjects: a population of 16 Japanese young adults with average age = 24.7 years old (4 women) and a population of 20 Japanese elderly people with average age = 65.15 years old (10 women). During the whole experiment, the heartbeat of each participant has been collected though Photopletysmography (PPG) using the Shimmer3 GSR+ Unit (www.shimmersensing.com) with a sampling frequency of 128 Hz. In addition to  PPG, the ElettroDermal Activity (EDA) of each subject has been acquired using the same sensor. The Shimmer3 GSR+ Units are non-invasive and completely painless sensors that could be easily worn by the participants as shown in Figure \ref{Fig:sensors}.

CLAWDAS data acquired during cognitive and listening tasks are here considered. 
The experimental protocol is composed of the following steps:
\begin{itemize}
    \item \textit{3 minutes} of questionnaires to collect the personal details and the current emotional state of each participant using \textbf{STAI Questionaries}.
    \item \textit{1 minute} of \textbf{Baseline} (BL) in resting state condition.
    \item \textit{6 minutes} of \textbf{Reading} (R) and \textbf{Comprehension} (C) tasks composed by two repetitions (trials) of 2 minutes of R followed by 1 minute of self assessment and C questions. 
    \item \textit{1 minute} of \textbf{Baseline} (BL) in resting state condition.
    \item A \textit{15 minutes} sequence composed of six repetitions of the following two tasks:
    \begin{enumerate}
        \item \textit{2 minutes} of \textbf{Audio Listening} (AL). In this task, relaxation is induced by natural sounds (Figure \ref{Fig:example_acquisition} right).
        \item \textit{30 seconds} of cognitive load, induced by mental \textbf{Math Calculations} (MC) that involve sums, subtractions and multiplications (Figure \ref{Fig:example_acquisition} left).
        \end{enumerate}
    Each repetition has a different audio track and math calculation.
    \item \textit{1 minute} of \textbf{Baseline} (BL) in resting state condition
\end{itemize}
 The experimental protocol has been reviewed and approved by the Research Ethics Committee at The University of Tokyo, Japan (No. 19-283 and 19-376). The CLAWDAS dataset is partitioned into two distinct subsets, according to the age of the participants: \textit{CLAWDAS Young}, that includes all the signals acquired from young adults and the \textit{CLAWDAS Elderly}, that groups all the signals collected from the elderly. In the following analysis the two groups are considered separately. 

Table \ref{tab:cardinality} reports the number of instances for each task in the CLAS and CLAWDAS datasets, keeping distinct the two subsets of CLAWDAS related to subjects' age (Young and Elderly).

\begin{table}[!t]
  \centering
  \caption{Number of instances for each task in the CLAS dataset (first 5 columns) and in the CLAWDAS dataset (last 5 columns), distinguished in CLAWDAS Young and CLAWDAS Elderly.  BL  =  Baseline, MP = Math Problems, ST = Stroop Test, LP = Logic Problems, NS = Neutral State, MC = Math Calculation, R = Reading, C = Comprehension.}
  \renewcommand{\arraystretch}{1.5}
  \begin{adjustwidth}{0.5cm}{0cm}
    \begin{tabular}{|c|c|ccccc||ccccc|}
    \hline  
    \multicolumn{1}{|c|}{} & \multicolumn{1}{p{1.8cm}|}{\textbf{Num Subj.}} & \multicolumn{1}{p{1.055em}}{\textbf{BL}} & 
    \multicolumn{1}{p{1.055em}}{\textbf{MP}} & 
    \multicolumn{1}{p{1.055em}}{\textbf{ST}} & 
    \multicolumn{1}{p{1.055em}}{\textbf{LP}} & 
    \multicolumn{1}{p{1.055em}||}{\textbf{NS}} & 
    \multicolumn{1}{p{1.055em}}{\textbf{BL}} & 
    \multicolumn{1}{p{1.78em}}{\textbf{MC}} & 
    \multicolumn{1}{p{1.055em}}{\textbf{AL}} & 
    \multicolumn{1}{p{1.055em}}{\textbf{R}} & 
    \multicolumn{1}{p{1.055em}|}{\textbf{C}}\\
    \hline \hline  
     \multicolumn{1}{|p{5em}|}{CLAS} & 60 & 60 & 60 & 60 & 60 & 360  & -  & -  & -  & -  & -  \\
    \hline  \hline  
    \multicolumn{1}{|p{5em}|}{CLAWDAS Young} & 16 & -  & -  & -  & -  & - &  46 & 96 & 96 & 32 & 32 \\
    \midrule  
    \multicolumn{1}{|p{5em}|}{CLAWDAS Elderly} & 20  & -  & -  & -  & - & - & 60 & 120 & 120 & 40 & 40\\
    \bottomrule
    \end{tabular}%
    \end{adjustwidth}
    \renewcommand{\arraystretch}{1}
  \label{tab:cardinality}%
\end{table}%

\section{Signal preprocessing: denosing and amplitude normalization}
\label{sec:preprocessing}
In this section the preprocessing operations applied to raw PPG data are detailed and differentiated for CLAS and CLAWDAS respectively.

\subsection{Denoising strategies}
The raw PPG signals are usually corrupted by noise and motion artifacts that can undermine their interpretation and use \cite{biswas2019motion}.  
In the CLAS dataset, the signals have been already preprocessed by the authors during the acquisition phase (\cite{markova2019clas}) and thus no further denosing procedure is applied. 

Concerning the CLAWDAS, the PPG raw signals of each subject are  preprocessed by a multiresolution wavelet denoising strategy, as suggested by \cite{biswas2019motion, raghuram2010performance}. The signal is divided in frequency sub-bands using Stationary Wavelet Transform (SWT) \cite{nason1995stationary} with mother wavelet \textit{Fejer-Korovkin} \cite{li2018wavelets}, and four levels of decomposition. A soft thresholding is applied to the detail coefficients of each sub-band. The universal threshold  calculated by the formula \(T_k = \sqrt{2log(N_j)}\) is adopted, where \textit{\(N_j\)} is the length of the $j-th$ wavelet coefficient, and \textit{k} is the sub-band, \cite{donoho1994ideal}. The SWT is implemented with the \textit{algorithm a-trous} \cite{holschneider1990real}. A preliminary operation of replicate padding is applied to the signal in order to obtain a length divisible by $2^{level}$ \cite{nason1995stationary}, with \textit{level} $=4$.

\subsection{Amplitude normalization}
\label{sec:AmpN}
In order to normalize the signals with respect to the amplitude, a Z-score operation, defined by the formula $Z=(x-\mu)/ \sigma$, is applied on the PPG recordings after the denoising procedure. 

In CLAWDAS the amplitude normalization, (\textbf{AmpN}), as well as the denoising preprocessing, is applied to the signal of each subject, before splitting it into the different experimental trials. 

A similar procedure is also applied to the CLAS dataset signals. However, in this case, the authors have already split the data into single trials, according to  their experimental protocol (see Section \ref{sec:dataset}), with no preliminary amplitude normalization. Thus, in order to apply a similar procedure to both the datasets, the segmented trials  of each subject are concatenated to re-build the original acquired signal.
Then, the Z-score amplitude normalization is applied to each subject signal. Finally, the amplitude normalized signals are split back into the trials, related to single tasks, using the markers properly defined during the previous phase of concatenation. 
\section{Personalized PPG normalization based on subject resting state heartbeat}
\label{sec:Normalization}

\begin{figure*}[!t]
    \centerline{\includegraphics[width=1.2\linewidth]{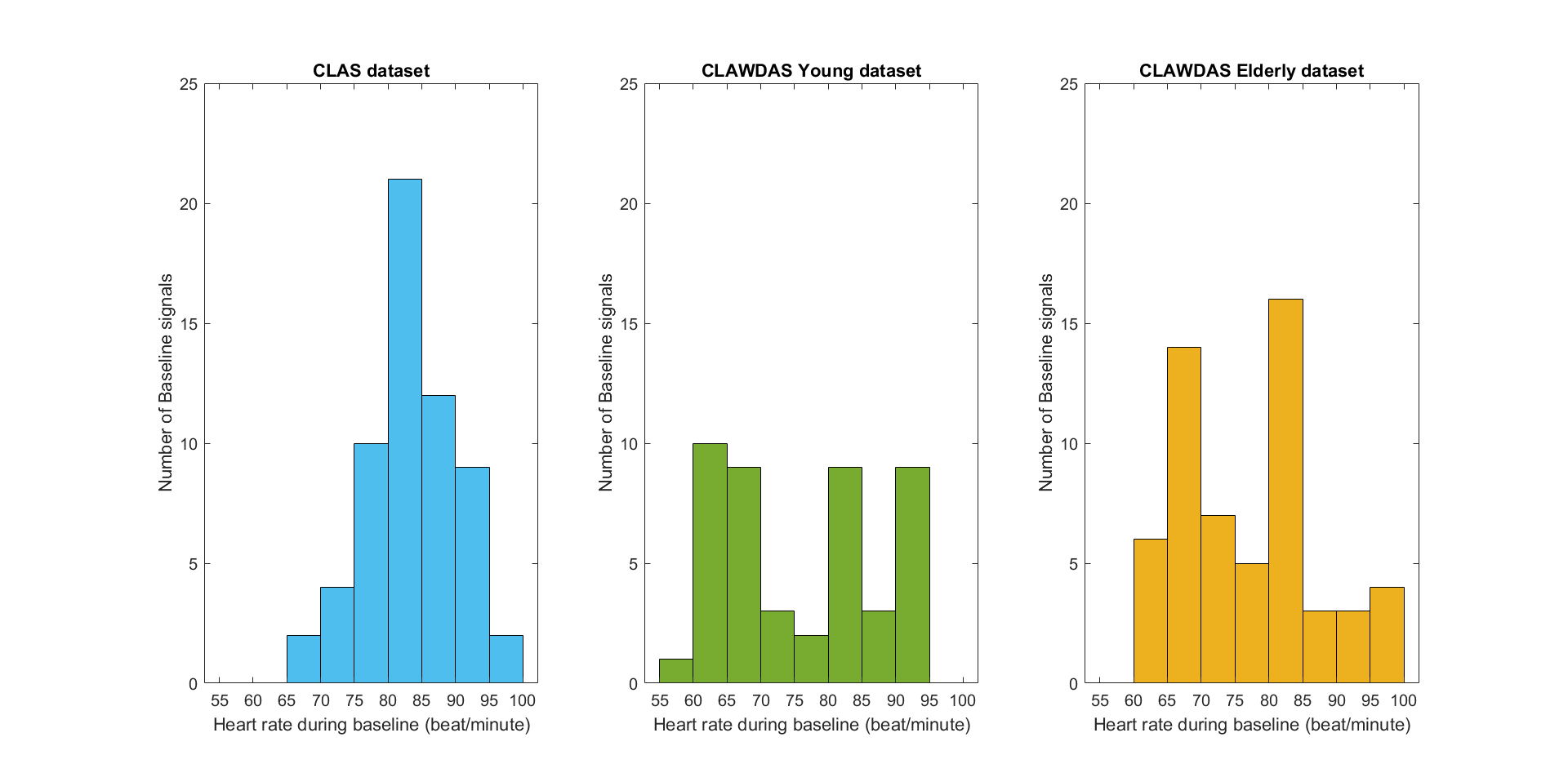}}
	\caption{Distributions of the average heartbeat of the resting state PPG signals in CLAS (left), CLAWDAS Young (middle) and CLAWDAS Elderly (right) datasets.}
	\label{fig:distribution}
	\hspace{0em}
\end{figure*}

The American Heart Association has underlined that the heart rate frequency of an adult in a resting state can vary in a range between 60 and 100 beats per minute. This inter-subject variability depends on many different factors both personal and environmental.
In case of CLAS and CLAWDAS datasets, the subjects' heartbeat range of the baseline recordings belongs to the one reported by the literature, as depicted in Figure \ref{fig:distribution}, where the average heartbeat distribution for each dataset is plotted. 

In order to get rid of this inter-subject variability, the core idea of our normalization procedure is to map PPG signals, defined in the Discrete Time Domain (DTD), into a new Subject Normalized discrete Domain (SND), applying a mapping procedure based on the resting state heartbeat frequency.
In this SND, all the subjects have the same resting state heart frequency.
For each subject, a personal resampling frequency can be calculated, so that for all the subjects the heart frequency of the resting state in the SND is equal, despite the original subject-peculiar frequency in the Continuous Time Domain (CTD). 

Then, the PPG data acquired during all the experimental tasks are also mapped into this new domain, applying the calculated personal resampling frequency and obtaining subject normalized PPG signals that can be considered for population-based analysis. 

Defining as ${f_c}[\frac{sample}{second}]$ the sampling frequency of the PPG signal, given by the acquisition device, and $f_{b}[\frac{beat}{second}]$ the heartbeat frequency in the resting state condition in CTD, the corresponding normalized heartbeat frequency $f_{Nb}[\frac{beat}{sample}]$ in the DTD is: 
\newline
\begin{linenomath}
\begin{equation}
\textit{$f_{Nb}$} = {\frac{f_b \frac{\textit{beat}}{\textit{second}}}{f_c\frac{\textit{sample}}{\textit{second}}}
	\label{eq:formulafNb}}
\end{equation}
\end{linenomath}
\newline
\begin{table}[!tb]
  \centering
  \caption{Correspondences between the three domains}
  \resizebox{\columnwidth}{!}{%
    \begin{tabular}{|c|c|c|c|}
    \hline
    \textbf{Domain} &
   Continuous Time & Discrete Time & Subject Normalized \\
    \hline
    \textbf{Acronym} &
   CTD & DTD & SND \\
   \hline
    \textbf {Heartbeat} &
    $f$ [$\frac{beat}{second}$] & 
    $f_{N}$ [$ \frac{beat}{sample}$]& 
    $f_{SN}$ [$\frac{beat}{SNsample}$] \\
    \hline
    \textbf {Resting state Heartbeat} &
    $f_{b}$ [$\frac{beat}{second}$] & 
    $f_{Nb}$ [$ \frac{beat}{sample}$]& 
    $f_{SNb}$ [$\frac{beat}{SNsample}$] \\
     \hline
   \textbf {Sampling frequency} & -- & $f_{c}$ [$\frac{sample}{second}$] & 
    $f_{SNc}$ [$\frac{SNsample}{sample}$] \\
     \hline
    \end{tabular}
    }
  \label{tab:frequencies}
\end{table}
We now define the Subject Normalized heartbeat frequency of the resting state in the SND as $f_{SNb}[\frac{beat}{SNsample}]$, where SNsample stands for Subject Normalized sample, that is the independent variable of the SND. 

The personal resampling frequency that permits to map the DTD PPG signal into the SND one is defined as $f_{SNc}[\frac{SNsample}{sample}]$, and can be calculated as follows:
\newline
\begin{linenomath}
\begin{equation}
\textit{$f_{SNc}$} = {\frac{f_{Nb} \frac{\textit{beat}}{\textit{sample}}}{f_{SNb}\frac{\textit{beat}}{\textit{SNsample}}}
	\label{eq:formulafSNc}}
\end{equation}
\end{linenomath}
\newline
As our goal is to obtain a domain where the inter-subject variability is discounted, the $f_{SNb}$ for all subjects' baseline should be the same. Once this constant value is chosen, the $f_{SNc}$ resampling frequency for each subject can be calculated from Eq. \ref{eq:formulafSNc}. Then, all the PPG data of the same subject can be resampled accordingly and mapped into the SND, making the SN-data reliable for population based analysis.
In Table \ref{tab:frequencies}, the notation introduced is summarized for the sake of clarity. 
Taking into account Eq. \ref{eq:formulafNb}, Eq. \ref{eq:formulafSNc} can be rewritten as follows:  
\newline
\begin{linenomath}
\begin{equation}
\textit{$f_{SNc}$} = {\frac{f_{b}}{f_{c}}}*{\frac{1}{f_{SNb}}} 
	\label{eq:formulafSNc2}
\end{equation}
\end{linenomath}
\newline
The resting state heartbeat in the SND ($f_{SNb}$) can be arbitrarily chosen, only paying attention to possible aliasing effects.  
In our calculations we set $f_{SNb}=\frac{1}{128} [\frac{beat}{SNsample}]$, that corresponds to one beat on 128 SNsamples in the SND.

To analyze the effect of our normalization proposal, let's consider some numerical examples.
In the case of a sampling frequency $f_{c}=128$ Hz as in the case of the CLAWDAS dataset, we can observe from Eq.\ref{eq:formulafSNc2} that, in the case of a subject with a baseline heartbeat frequency of $60 \frac{beat}{minute}$, corresponding to $1  \frac{beat}{second}$, the personalized resampling frequency is $f_{SNc}=1 \frac{SNsample}{sample}$, meaning that there are no differences between the signal in the DTD and in the SND. Note that $60 \frac{beat}{minute}$ is generally considered as the minimum value for normal people.   
For heartbeat frequencies higher than $60 \frac{beat}{minute}$ the mapping from DTD to SND implies an over-sampling, while for lower frequencies, the consequent under-sampling does not introduce aliasing as 128 samples are guaranteed between two consecutive peaks.

\begin{figure*}[!t]
    \centerline{\includegraphics[width=1.2\linewidth]{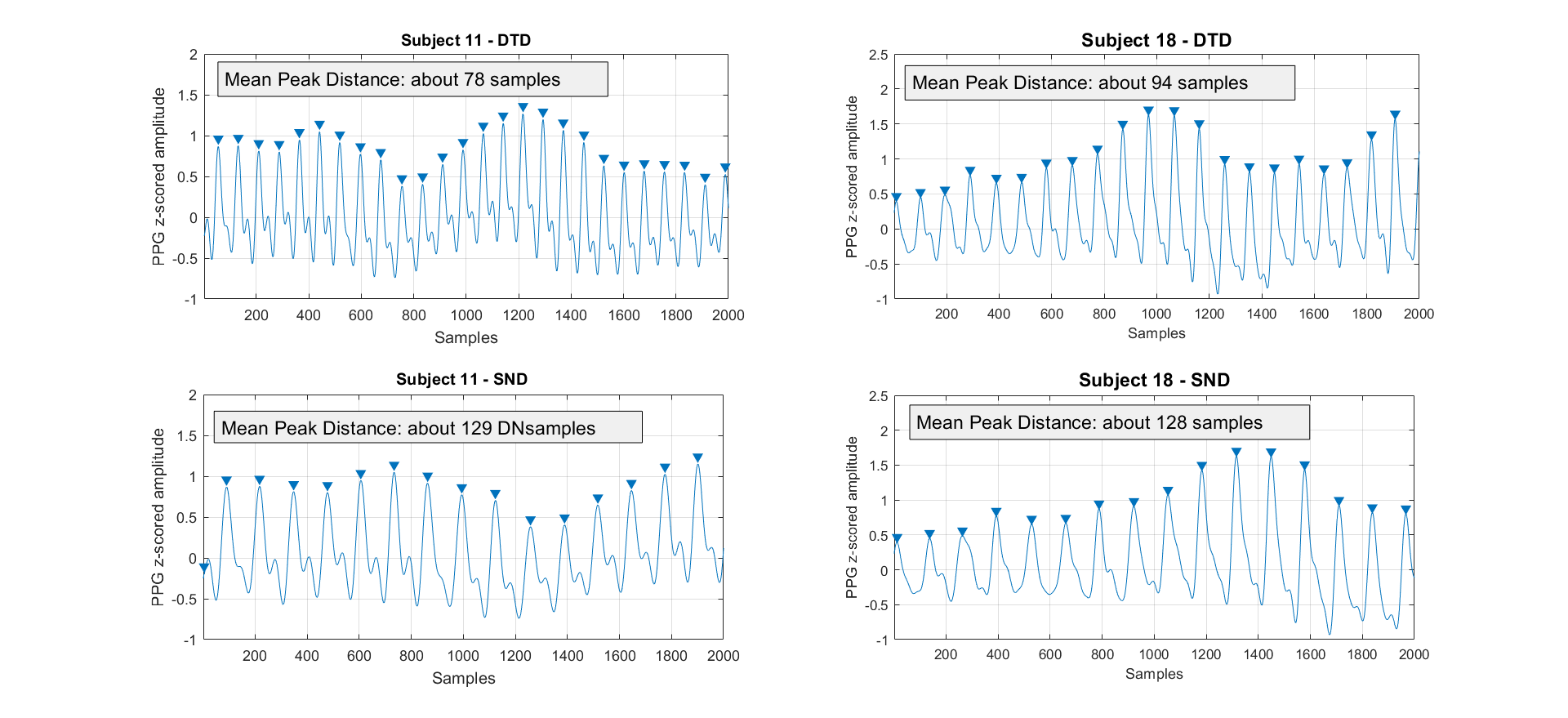}}
	\caption{PPG data in resting state conditions for subjects 11 (left) and 18 (right) are here reported, in the original discrete time domain (DTD) and in the Subject-Normalized Domain (SND), (top and bottom rows respectively). 
	Before the proposed normalization (top row), the subjects have different heartbeat frequencies. While, in the SND they are more similar.}
	\label{fig:norm_signal}
	\hspace{0em}
\end{figure*}

In Figure \ref{fig:norm_signal}, PPG signals corresponding to the first baseline in the CLAWDAS Elderly dataset of subjects 11 and 18 respectively are considered. In the first row, the signals in the DTD are reported, showing the difference between the two subjects' heartbeat frequency: for subject 11   $f_{Nb}=\frac{1}{78}\frac{beat}{sample}$, corresponding to $84 \frac{beat}{minute}$, while for subject 18 is  $f_{Nb}=\frac{1}{94}\frac{beat}{sample}$ corresponding to about $82 \frac{beat}{minute}$.
In the second row the same two signals resampled in the SND are shown. Note that we assumed in defining our procedure that the heartbeat during a resting state is a stationary and periodic signal, however this is not the case in real life, justifying not having $f_{Nb}$ strictly equal to $\frac{1}{128} \frac{beat}{SNsample}$ for both the subjects in the SND. 

In case of multiple baseline signals, the heartbeat frequency is evaluate as the average of the heartbeat frequency of all of them. This procedure is applied, for example, during the normalization of the CLAWDAS signals. In this case, in fact, three different baseline signals have been acquired from each subject.

\section{Classification Setting}
\label{sec:classificationSetting}
A binary classification task is here proposed on CLAS, CLAWDAS Young and CLAWDAS Elderly, to evaluate the performance of the personalized PPG normalization presented in this work. 

Two classes are considered: the class corresponding to signals collected during high cognitive load tasks: \textit{High CL}, and the class related to low cognitive load tasks: \textit{Low CL} . The tasks used as representative of each class change according to the dataset considered.

In particular, according to \cite{markova2019clas}, in the CLAS dataset the PPG data collected during the three cognitive tasks (MP, ST and LP) are labeled as \textit{High CL}, while the NS data are labeled as \textit{Low CL}. In order to make the two classes balanced, MP, ST and LP signals are split into two non-overlapped segments of equal length, reaching a cardinality of 360 instances for both classes.

In the CLAWDAS Young and CLAWDAS Elderly, the data collected during the MC task are selected for the \textit{High CL} class, while the data collected during the AL task are chosen as representatives of the \textit{Low CL} class. Thereby, the two classes are equally balanced, with 96 instances each for the CLAWDAS Young, and 120 instances each for the CLAWDAS Elderly.

In all the performed analysis, seven handcrafted features are extracted as characteristics useful to describe the PPG signals:
\begin{itemize}
    \item \textbf{Minima}, \textbf{Maxima}, \textbf{Mean} and \textbf{Standard Deviation} of the signal
    \item \textbf{Peak Rate}, which represents the mean number of peaks
    \item \textbf{Inter Beat Interval (IBI)}, which represents the mean distance between two peaks in a row 
    \item \textbf{Root Mean Square of Successive Distance (RMSSD)} which represents the variance of the distance between two consecutive peaks \cite{stein1994heart}.
\end{itemize}
The last three features are evaluated in the discrete domain and reported with respect to \textit{samples}. For the sake of clarity, it is recalled that the meaning of \textit{samples} changes according to the type of normalization strategy adopted. In particular, \textit{samples} refer to subject normalized samples when the features is evaluated on signals with personalized normalization based on resting state heartbeat, while it refers to discrete time samples in all the other cases. 
All the features so evaluated are also standardized applying z-score before being used as input to the different classifiers.
%
\begin{table*}[!t]
  \centering
  \caption{Summary of the classification settings}
    \begin{adjustwidth}{-2.7cm}{0cm}
    \begin{tabular}{lllll}
    \toprule
    \multicolumn{1}{p{12em}}{\textbf{Types of Normalization considered}} & \multicolumn{4}{p{28.055em}}{AmpN, SubjFeatN, PersFreqN} \\
       &    &    &    &   \\
    \midrule
    \textbf{Features Used} & \multicolumn{4}{l}{Maximum, Minimum, Mean, Standard deviation, Peak Rate, IBI, RMSSD} \\
       &    &    &    &  \\
    \midrule
    \textbf{Classifier Involved} & \multicolumn{4}{l}{SVM Linear, SVM Cubic, SVM Gauss and Cart} \\
       &    &    &    &  \\
    \midrule
    \multicolumn{1}{p{12em}}{\textbf{Performance Evaluation method}} & LOSO &    &    &  \\
       &    &    &    &  \\
    \midrule
    \textbf{Evaluation Metrics} & \multicolumn{4}{l}{Accuracy, single class F1-score} \\
    &    &    &    &  \\
    \midrule
    \midrule
    \multicolumn{1}{{p{12em}}}{\textbf{Dataset}} & \multicolumn{2}{c|}{\textbf{CL High Class}} & \multicolumn{2}{c}{\textbf{CL Low Class}} \\
       & \multicolumn{1}{c}{\textbf{Task}} & \multicolumn{1}{c|}{\textbf{Num. of signals}} & \multicolumn{1}{c}{\textbf{Task}} & \multicolumn{1}{c}{\textbf{Num. of signals}} \\
    \midrule
    CLAS & \multicolumn{1}{p{7.72em}}{Math, Stroop and Logic Test} & \multicolumn{1}{c|}{360} & Neutral State & \multicolumn{1}{c}{360} \\
    CLAWDAS Young & Math Calculation & \multicolumn{1}{c|}{96} & Audio Listening & \multicolumn{1}{c}{96} \\
    CLAWDAS Elderly & Math Calculation & \multicolumn{1}{c|}{120} & Audio Listening & \multicolumn{1}{c}{120} \\
    \bottomrule
    \end{tabular}%
    \end{adjustwidth}
  \label{tab:summary_class_setting}%
\end{table*}%

Using the the seven features above introduced, three binary classification experiments are performed for each of the two datasets considered, comparing the following three normalization strategies: 
\begin{itemize}
    \item \textbf{AmpN}: Amplitude normalization, as described in \ref{sec:AmpN};
    \item \textbf{SubjFeatN}: Amplitude normalization followed by a subject feature normalization. This feature normalization is performed with respect to the subject baseline on peak rate, IBI and RMSSD features as follows: 
    \newline
    \begin{linenomath}
    \begin{equation}
		featNorm_i = \frac{feature_i - \overline{featureBL_i}}{\overline{featureBL_i}}
	\end{equation}
	\end{linenomath}
	\newline
    where $i \in {Peak Rate, IBI, RMSSD}$; $feature_i$ represents the feature value before the normalization; $featNorm_i$ the new normalized value; and $\overline{featureBL_i}$ the mean value of the $i-th$ feature evaluated on the subject resting state;  
    \item \textbf{PersFreqN}: Amplitude normalization followed by the personalized normalization based on resting state heartbeat, described in chapter \ref{sec:Normalization}.
\end{itemize}
\  \\
For each analysis, four different classification models are tested: a Classification and Regression Tree \textit{(Cart)} with Gini’s diversity index as criterion of splitting and 100 as max number of decision split, and three Support Vector Machine \textit{(SVM)} with different kernels:  Linear \textit{(SVM Linear)}, Gaussian \textit{(SVM Gaussian)} and polynomial cubic \textit{(SVM Cubic)}. In particular, for the gaussian kernel SVM, the kernel scale is set to 3.3 in order to consider a Medium Gaussian SVM. 

A Leave One Subject Out (\textit{LOSO}) Cross Validation (\cite{schmidt2019wearable}) is applied to evaluate the performance of the trained classifiers. At each iteration, the data used to train the classifier consists of the signals collected from all the subjects except one, whose instances are instead used to test the performance of the model. 
An overall confusion matrix is finally generated, joining the single confusion matrices resulting from each iteration. From this confusion matrix, several well-known evaluation metrics are extracted. In particular, we have selected the \textit{accuracy} to evaluate the general performance of the classifier and the single class \textit{ F1-score} (\cite{bishop2006pattern}) to assess, instead, the goodness of the classification model in recognizing the single classes.
The classification settings described above are summarized in Table \ref{tab:summary_class_setting}.


\section{Results and discussion}
The performance of the classification settings described in Table \ref{tab:summary_class_setting}  are reported in Tables \ref{tab:performance CLAS}, \ref{tab:performance_japan_yng} and \ref{tab:performance_japan_eld}, for CLAS, CLAWDAS Young and CLAWDAS Elderly, datasets respectively.
In particular, the results of the different normalization strategies are reported in term of \textit{accuracy} and single class \textit{F1-score} generated using the LOSO Cross Validation approach. 

\begin{figure}[!t]
\centerline{\includegraphics[width=1.3 \columnwidth]{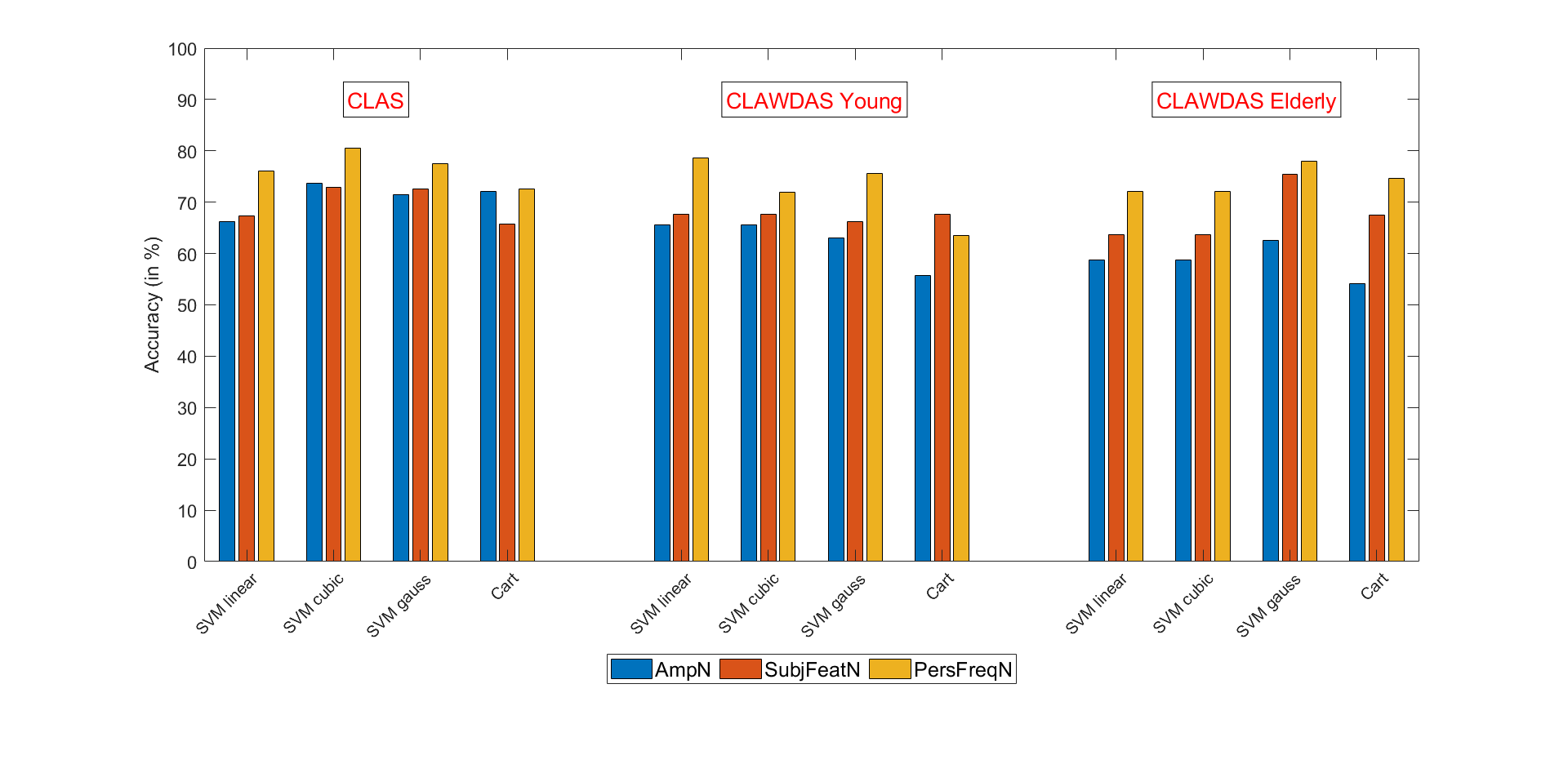}}
\caption{Bar plot comparison of the accuracy obtained using different classifiers and normalization procedures. Three dataset are considered: CLAS, CLAWDAS Young and CLAWDAS Elderly}
\label{fig:Accuracy_BarPlot}
\end{figure}

In all the experiments carried out, the normalization strategy here proposed, \textbf{PersFreqN}, outperforms the other two normalization procedures, for all the datasets. This observation is further supported by a visual comparison of the performance of the classifiers reported in the bar plot of Figure \ref{fig:Accuracy_BarPlot}, varying the normalization strategy and the involved dataset.

Comparing the datasets, the best performance is observed on the CLAS dataset. In this case, the proposed \textbf{PersFreqN} allows to reach an accuracy of 81\% adopting the SVM classifier with polynomial cubic kernel. This result significantly outperforms the accuracy of 74\% reached using \textbf{AmpN} and 73\% obtained using \textbf{SubjFeatN}.

In the CLAWDAS datasets, the highest accuracy achieved are 79\% for CLAWDAS Young and 80\% for CLAWDAS Elderly, both obtained with the proposed \textbf{PersFreqN} and SVM with linear kernel. These values show a significant improvement with respect to the other normalization strategies \textbf{AmpN} and \textbf{SubjFeatN}, that are always lower than 66\%. \newline

\begin{table*}[!tbp]
  \centering
  \renewcommand{\arraystretch}{1.1}
  \caption{Performance comparison on CLAS dataset, varying the  normalization strategies (columns) and classification models (rows). Two performance metrics are evaluated using a LOSO validation strategy: Accuracy (\textit{Acc}) and single class \textit{F1-Score}. The best performances reached for each type of normalization are underlined, while the the highest accuracy value of all is highlighted in bold.}
  \begin{adjustwidth}{-2cm}{0cm}
    \begin{tabular}{|c|ccc|ccc|ccc|}
      \hline
    & \multicolumn{3}{c|}{\textbf{AmpN}} & \multicolumn{3}{c|}{\textbf{SubjFeatN}} & \multicolumn{3}{c|}{\textbf{PersFreqN}} \\
    \hline
    \multicolumn{1}{|c|}{\multirow{2}{*}{\textbf{Classifier}}} & 
    \multicolumn{1}{p{1em}}{} & 
    \multicolumn{1}{p{3.6em}}{\centering \small{\textbf{High CL}}} & 
    \multicolumn{1}{p{3.6em}|}{\centering \small{\textbf{Low CL}}} & 
    \multicolumn{1}{p{1em}}{} & 
    \multicolumn{1}{p{3.6em}}{\centering \small{\textbf{High CL}}} & 
    \multicolumn{1}{p{3.6em}|}{\centering \small{\textbf{Low CL }}} & 
    \multicolumn{1}{p{1em}}{} & 
    \multicolumn{1}{p{3.6em}}{\centering \small{\textbf{High CL}}} & 
    \multicolumn{1}{p{3.6em}|}{\centering \small{\textbf{Low CL}}} \\
    & 
    \multicolumn{1}{p{1em}}{\centering \textbf{Acc}}& 
    \multicolumn{1}{p{3.6em}}{\centering \small{\textbf{F1-Score}}} & 
    \multicolumn{1}{p{3.6em}|}{\centering \small{\textbf{F1-Score}}} & 
    \multicolumn{1}{p{1em}}{\centering \textbf{Acc}} & 
    \multicolumn{1}{p{3.6em}}{\centering \small{\textbf{F1-Score}}} & 
    \multicolumn{1}{p{3.6em}|}{\centering \small{\textbf{F1-Score}}} & 
    \multicolumn{1}{p{1em}}{\centering \textbf{Acc}} & 
    \multicolumn{1}{p{3.6em}}{\centering \small{\textbf{F1-Score}}} & 
    \multicolumn{1}{p{3.6em}|}{\centering \small{\textbf{F1-Score}}} \\
    \hline
       &    &    &    &    &    &    &    &    &  \\
    SVM Linear & \multicolumn{1}{c}{66\%} & \multicolumn{1}{c}{0,65} & \multicolumn{1}{c|}{0,67} & \multicolumn{1}{c}{67\%} & \multicolumn{1}{c}{0,68} & \multicolumn{1}{c|}{0,67} & \multicolumn{1}{c}{76\%} & \multicolumn{1}{c}{0,77} & \multicolumn{1}{c|}{0,76} \\
    SVM Cubic & \multicolumn{1}{c}{\underline{74\%}} & \multicolumn{1}{c}{0,74} & \multicolumn{1}{c|}{0,73} & \multicolumn{1}{c}{\underline{73\%}} & \multicolumn{1}{c}{0,74} & \multicolumn{1}{c|}{0,72} & \multicolumn{1}{c}{{\textbf{\underline{81\%}}}} & \multicolumn{1}{c}{0,81} & \multicolumn{1}{c|}{0,81} \\
    SVM Gauss & \multicolumn{1}{c}{72\%} & \multicolumn{1}{c}{0,72} & \multicolumn{1}{c|}{0,71} & \multicolumn{1}{c}{73\%} & \multicolumn{1}{c}{0,72} & \multicolumn{1}{c|}{0,73} & \multicolumn{1}{c}{78\%} & \multicolumn{1}{c}{0,78} & \multicolumn{1}{c|}{0,77} \\
    Cart & \multicolumn{1}{c}{72\%} & \multicolumn{1}{c}{0,71} & \multicolumn{1}{c|}{0,73} & \multicolumn{1}{c}{66\%} & \multicolumn{1}{c}{0,65} & \multicolumn{1}{c|}{0,66} & \multicolumn{1}{c}{73\%} & \multicolumn{1}{c}{0,72} & \multicolumn{1}{c|}{0,73} \\
       &    &    &    &    &    &    &    &    &  \\
    \hline
    \end{tabular}%
  \end{adjustwidth}
  \renewcommand{\arraystretch}{1}
  \label{tab:performance CLAS}%
\end{table*}%

\begin{table*}[!t]
  \centering
  \caption{Performance comparison on CLAWDAS Young dataset, varying the  normalization strategies (columns) and classification models (rows). Two performance metrics are evaluated using a LOSO validation strategy: Accuracy (\textit{Acc}) and single class \textit{F1-Score}. The best performances reached for each type of normalization are underlined, while the the highest accuracy value of all is highlighted in bold.}
    \renewcommand{\arraystretch}{1.1}
    \begin{adjustwidth}{-2cm}{0cm}
    \begin{tabular}{|c|ccc|ccc|ccc|}
    \hline
    & \multicolumn{3}{c|}{\textbf{AmpN}} & \multicolumn{3}{c|}{\textbf{SubjFeatN}} & \multicolumn{3}{c|}{\textbf{PersFreqN}} \\
    \hline
    \multicolumn{1}{|c|}{\multirow{2}{*}{\textbf{Classifier}}} & 
    \multicolumn{1}{p{1em}}{} & 
    \multicolumn{1}{p{3.6em}}{\centering \small{\textbf{High CL}}} & 
    \multicolumn{1}{p{3.6em}|}{\centering \small{\textbf{Low CL}}} & 
    \multicolumn{1}{p{1em}}{} & 
    \multicolumn{1}{p{3.6em}}{\centering \small{\textbf{High CL}}} & 
    \multicolumn{1}{p{3.6em}|}{\centering \small{\textbf{Low CL }}} & 
    \multicolumn{1}{p{1em}}{} & 
    \multicolumn{1}{p{3.6em}}{\centering \small{\textbf{High CL}}} & 
    \multicolumn{1}{p{3.6em}|}{\centering \small{\textbf{Low CL}}} \\
    & 
    \multicolumn{1}{p{1em}}{\centering \textbf{Acc}}& 
    \multicolumn{1}{p{3.6em}}{\centering \small{\textbf{F1-Score}}} & 
    \multicolumn{1}{p{3.6em}|}{\centering \small{\textbf{F1-Score}}} & 
    \multicolumn{1}{p{1em}}{\centering \textbf{Acc}} & 
    \multicolumn{1}{p{3.6em}}{\centering \small{\textbf{F1-Score}}} & 
    \multicolumn{1}{p{3.6em}|}{\centering \small{\textbf{F1-Score}}} & 
    \multicolumn{1}{p{1em}}{\centering \textbf{Acc}} & 
    \multicolumn{1}{p{3.6em}}{\centering \small{\textbf{F1-Score}}} & 
    \multicolumn{1}{p{3.6em}|}{\centering \small{\textbf{F1-Score}}} \\
    \hline
     &  &    &    &    &    &    &    &    &  \\
    SVM Linear & \multicolumn{1}{c}{\underline{66\%}} & \multicolumn{1}{c}{0,62} & \multicolumn{1}{c|}{0,69} & \multicolumn{1}{c}{\underline{68\%}} & \multicolumn{1}{c}{0,63} & \multicolumn{1}{c|}{0,72} & \multicolumn{1}{c}{\textbf{\underline{79\%}}} & \multicolumn{1}{c}{0,77} & \multicolumn{1}{c|}{0,80} \\
    SVM Cubic & \multicolumn{1}{c}{\underline{66\%}} & \multicolumn{1}{c}{0,66} & \multicolumn{1}{c|}{0,66} & \multicolumn{1}{c}{\underline{68\%}} & \multicolumn{1}{c}{0,67} & \multicolumn{1}{c|}{0,68} & \multicolumn{1}{c}{72\%} & \multicolumn{1}{c}{0,72} & \multicolumn{1}{c|}{0,72} \\
    SVM Gauss & \multicolumn{1}{c}{63\%} & \multicolumn{1}{c}{0,59} & \multicolumn{1}{c|}{0,66} & \multicolumn{1}{c}{66\%} & \multicolumn{1}{c}{0,61} & \multicolumn{1}{c|}{0,70} & \multicolumn{1}{c}{76\%} & \multicolumn{1}{c}{0,75} & \multicolumn{1}{c|}{0,76} \\
    Cart & \multicolumn{1}{c}{56\%} & \multicolumn{1}{c}{0,59} & \multicolumn{1}{c|}{0,51} & \multicolumn{1}{c}{\underline{68\%}} & \multicolumn{1}{c}{0,67} & \multicolumn{1}{c|}{0,69} & \multicolumn{1}{c}{64\%} & \multicolumn{1}{c}{0,63} & \multicolumn{1}{c|}{0,64} \\
    &    &    &    &    &    &    &    &    &  \\
    \bottomrule
    \end{tabular}%
  \end{adjustwidth}
  \renewcommand{\arraystretch}{1}
  \label{tab:performance_japan_yng}%
\end{table*}%

\begin{table*}[!t]
  \centering
  \caption{Performance comparison on CLAWDAS Elderly, varying the  normalization strategies (columns) and classification models (rows). Two performance metrics are evaluated using a LOSO validation strategy: Accuracy (\textit{Acc}) and single class \textit{F1-Score}. The best performances reached for each type of normalization are underlined, while the the highest accuracy value of all is highlighted in bold.}
  \renewcommand{\arraystretch}{1.1}
  \begin{adjustwidth}{-2cm}{0cm}
    \begin{tabular}{|c|ccc|ccc|ccc|}
      \hline
    & \multicolumn{3}{c|}{\textbf{AmpN}} & \multicolumn{3}{c|}{\textbf{SubjFeatN}} & \multicolumn{3}{c|}{\textbf{PersFreqN}} \\
    \hline
    \multicolumn{1}{|c|}{\multirow{2}{*}{\textbf{Classifier}}} & 
    \multicolumn{1}{p{1em}}{} & 
    \multicolumn{1}{p{3.6em}}{\centering \small{\textbf{High CL}}} & 
    \multicolumn{1}{p{3.6em}|}{\centering \small{\textbf{Low CL}}} & 
    \multicolumn{1}{p{1em}}{} & 
    \multicolumn{1}{p{3.6em}}{\centering \small{\textbf{High CL}}} & 
    \multicolumn{1}{p{3.6em}|}{\centering \small{\textbf{Low CL }}} & 
    \multicolumn{1}{p{1em}}{} & 
    \multicolumn{1}{p{3.6em}}{\centering \small{\textbf{High CL}}} & 
    \multicolumn{1}{p{3.6em}|}{\centering \small{\textbf{Low CL}}} \\
    & 
    \multicolumn{1}{p{1em}}{\centering \textbf{Acc}}& 
    \multicolumn{1}{p{3.6em}}{\centering \small{\textbf{F1-Score}}} & 
    \multicolumn{1}{p{3.6em}|}{\centering \small{\textbf{F1-Score}}} & 
    \multicolumn{1}{p{1em}}{\centering \textbf{Acc}} & 
    \multicolumn{1}{p{3.6em}}{\centering \small{\textbf{F1-Score}}} & 
    \multicolumn{1}{p{3.6em}|}{\centering \small{\textbf{F1-Score}}} & 
    \multicolumn{1}{p{1em}}{\centering \textbf{Acc}} & 
    \multicolumn{1}{p{3.6em}}{\centering \small{\textbf{F1-Score}}} & 
    \multicolumn{1}{p{3.6em}|}{\centering \small{\textbf{F1-Score}}} \\
    \hline
     &  &    &    &    &    &    &    &    &  \\
    SVM Linear & \multicolumn{1}{c}{59\%} & \multicolumn{1}{c}{0,58} & \multicolumn{1}{c|}{0,60} & \multicolumn{1}{c}{69\%} & \multicolumn{1}{c}{0,65} & \multicolumn{1}{c|}{0,72} & \multicolumn{1}{c}{\underline{\textbf{80\%}}} & \multicolumn{1}{c}{0,80} & \multicolumn{1}{c|}{0,81} \\
    SVM Cubic & \multicolumn{1}{c}{59\%} & \multicolumn{1}{c}{0,55} & \multicolumn{1}{c|}{0,62} & \multicolumn{1}{c}{64\%} & \multicolumn{1}{c}{0,61} & \multicolumn{1}{c|}{0,66} & \multicolumn{1}{c}{72\%} & \multicolumn{1}{c}{0,71} & \multicolumn{1}{c|}{0,73} \\
    SVM Gauss & \multicolumn{1}{c}{\underline{63\%}} & \multicolumn{1}{c}{0,61} & \multicolumn{1}{c|}{0,64} & \multicolumn{1}{c}{\underline{75\%}} & \multicolumn{1}{c}{0,74} & \multicolumn{1}{c|}{0,76} & \multicolumn{1}{c}{78\%} & \multicolumn{1}{c}{0,78} & \multicolumn{1}{c|}{0,78} \\
    Cart & \multicolumn{1}{c}{54\%} & \multicolumn{1}{c}{0,51} & \multicolumn{1}{c|}{0,57} & \multicolumn{1}{c}{68\%} & \multicolumn{1}{c}{0,68} & \multicolumn{1}{c|}{0,67} & \multicolumn{1}{c}{75\%} & \multicolumn{1}{c}{0,75} & \multicolumn{1}{c|}{0,74} \\
        &    &    &    &    &    &    &    &    &  \\
    \bottomrule
    \end{tabular}%
  \end{adjustwidth}
  \renewcommand{\arraystretch}{1}
  \label{tab:performance_japan_eld}%
\end{table*}%

Another consideration regards the classifier that allows to reach the best results. In general, from the three Tables  \ref{tab:performance CLAS}, \ref{tab:performance_japan_yng} and \ref{tab:performance_japan_eld} it emerges that the highest accuracy values are usually achieved by the SVM classifiers with linear or cubic kernel, whereas the lowest ones are generally obtained by the Cart classifier.

Finally, a last consideration should be done on the \textbf{SubjFeatN} normalization strategy. 
In general, this normalization produces performance higher than the \textbf{AmpN} one, even if it appears less effective compared to the proposed \textbf{PersFreqN} strategy. These results confirm that a normalization strategy that takes into account not only amplitude normalization but also subject's characteristics, should be adopted to remove inter-subject variability.    




As a final remark, all the adopted classification settings are able to classify with comparable performance both the two classes, as indicated by the values of the single class \textit{F1-score} in all the Tables. However, the introduction of the proposed \textbf{PersFreqN} seems to produced even more balanced classification results.

\label{sec:results}
\section{Conclusions}
While considering physiological data, signal normalization  not only with respect to amplitude but also with respect to personal characteristics is mandatory to perform subject based analysis, especially if machine learning techniques should be applied.  Personalized normalization on PPG data, both at the feature level, \textbf{SubjFeatN},  and with respect to heartbeat frequency, \textbf{PersFreqN}, introduces an increase in the classification performance, considering different classification models and datasets. In particular, the personalized PPG normalization based on subject heartbeat here proposed, \textbf{PersFreqN}, outperforms the other strategies and permits to significantly reduce inter-subject heterogeneity. 
Moreover, the proposed normalization could be also useful for intra-subject analysis, especially when comparing the physiological responses of the same subject, in different days or even in different moment of the day: it is well known, in fact, that the physiological responses not only depend on external stimuli, but also on physical and internal conditions that can significantly vary for the same subject with respect to time.    

\section*{Author contributions}
F.G: Conceptualisation, methodology, formal analysis, investigation, data curation, writing original draft preparation; A.G: software, validation, formal analysis, investigation, data curation, writing original draft preparation, visualisation; M.G: investigation, writing review and editing; S.B.: project leader, supervision.
All authors have read and agreed to the published version of the manuscript.

\section*{Acknowledgments} 
This research is partially supported by the FONDAZIONE CARIPLO “LONGEVICITY-Social Inclusion for the Elderly through Walkability” (Ref. 2017-0938) and by the Japan Society for the Promotion of Science (Ref. L19513). We want to give our thanks to Prof. Katsuhiro Nishinari and his staff, in particular Kenichiro Shimura and Daichi Yanagisawa for their indispensable support during the experiment  held  at  RCAST  -  The  University  of  Tokyo.

\abbreviations{Abbreviations}{
The following abbreviations are used in this manuscript:\\

\noindent 
\begin{tabular}{@{}ll}
PPG & Photopletysmograpy\\
\\
CLAS dataset & Cogntive Load, Affect and Stress Recognition dataset\\
CLAWDAS dataset & Cognitive Load and Affective Walkability in Different Age Subjects dataset\\
\\
BL  & Baseline\\
MP &  Math Problems\\
ST & Stroop Test\\ 
LP & Logic Problems\\
NS & Neutral State\\
MC & Math Calculation \\
R & Reading\\
C & Comprehension\\
\\
SWT & Stationary Wavelet Transform\\
\\
CTD & Continuous Time Domain\\
DTD & Discrete Time Domain\\
SND & Subject Normalized discrete Domain \\
\\
High CL & High Cogntive Load \\
Low CL & Low Cogntive Load \\
\\
IBI & Inter Beat Ineterval \\
RMSSD & Root Mean of Successive Distance\\
\\
SVM & Support Vector Machine\\
SVM Linear & Support Vector Machine with Linear Kernel\\
SVM Cubic & Support Vector Machine with Polynomial Cubic Kernel\\
SVM Gaus & Support Vector Machine with Gaussian Kernel\\
Cart & Classification and Regression Tree\\
\\
LOSO Cross Validation & Leave One Subject Out Cross Validation\\
\\
AmpN & Amplitude Normalization\\
SubjFeatN & Amplitude normalization followed by a Subject Feature Normalization.\\
PersFreqN & Personalized Normalization based on baseline heartbeat Frequency
\end{tabular}}

\newpage

\begin{adjustwidth}{-\extralength}{0cm}

\reftitle{References}


\bibliography{bibliography}


\end{adjustwidth}

\end{document}